\documentclass[aip,twocolumn,showpacs,amsmath,amssymb,superscriptaddress]{revtex4}
\usepackage{graphicx}
\usepackage{subfigure}
\usepackage{grffile}
\usepackage{bm}
\usepackage{color}
\usepackage{hyperref}
\newcommand{\bw}{\begin{widetext}}
\newcommand{\ew}{\end{widetext}}
\newcommand{\be}{\begin{equation}}
\newcommand{\ee}{\end{equation}}
\newcommand{\ba}{\begin{eqnarray}}
\newcommand{\ea}{\end{eqnarray}}

\newcommand{\1}{\mbox{\bf 1}}

\begin{document}
\title{Shape transition of unstrained flattest single-walled carbon nanotubes under pressure}

\author{Weihua Mu}
\email{whmu@mit.edu, muwh@itp.ac.cn}
\affiliation{Department of Chemistry, Massachusettes Institute of Technology, Cambridge, Massachusetts 02139, United States}
\affiliation{State Key Laboratory of Theoretical Physics, Institute of Theoretical Physics, The Chinese Academy of
Sciences, Kavli Institute for Theoretical Physics China, The Chinese Academy of Sciences, P. O. Box 2735 Beijing 100190, China}

\author{Jianshu Cao}
\email{jianshu@mit.edu}
\affiliation{Department of Chemistry, Massachusettes Institute of Technology, Cambridge, Massachusetts 02139, United States}
\affiliation{Singapore-MIT Alliance for Research and Technology (SMART), Singapore, 138602}

\author{Zhong-can Ou-Yang}
\affiliation{State Key Laboratory of Theoretical Physics, Institute of Theoretical Physics, The Chinese Academy of
Sciences, Kavli Institute for Theoretical Physics China, The Chinese Academy of Sciences, P. O. Box 2735 Beijing 100190, China}
\affiliation{Center for Advanced Study, Tsinghua University, Beijing 100084, China}

\date{\today}

\begin{abstract}
Single walled carbon nanotube's cross section can be flattened under hydrostatic pressure. One example is the cross section of a single walled carbon nanotube successively deforms from the original round shape to oval shape, then to peanut-like shape. At the transition point of reversible deformation between convex shape and concave shape, the side wall of nanotube is flattest. This flattest tube has many attractive properties. In the present work, an approximate approach is developed to determine the equilibrium shape of this unstrained flattest tube and the curvature distribution of this tube. Our results are in good agreement with recent numerical results, and can be applied to the study of pressure controlled electric properties of single walled carbon nanotubes. The present method can also be used to study other deformed inorganic and organic tube-like structures.
\end{abstract}
\maketitle
Carbon nanotubes (CNTs) have extraordinary mechanical properties due to
strong carbon-carbon atomic interactions in their honeycomb lattices.~\cite{saito1998} CNTs' high elastic modulus, exceptional axial stiffness, and low density, make them ideal for nanotechnology applications.~\cite{jorio2008}. CNTs' mechanical properties are highly anisotropic, and have been studied extensively, e.g. Ref.~[\onlinecite{tang2000,anis2012}] and the references therein. In contrast to the high tensile strength~\cite{treacy1996}, CNTs are susceptible to mechanical distortion in their radial directions under applied hydrostatic pressure on the scale of several GPa.~\cite{yao2008,caillier2008,gao1998, aguiar2011,elliott2004,sluiter2004,tangney2005,imtani2007,yang2007b,zang2004,sun2004,cai2006}.
The radial deformation controlled by the applied pressure provides an approach to modify the electronic properties of SWCNTs~\cite{tang2002}, consequently, radial deformation of SWCNTs can be observed by optical spectroscopy since the electronic band structure of a SWCNT is sensitive to its morphological transition~\cite{anis2012,thirunavukkuarasu2010}.  In addition, a SWCNT's chemical reactivity depends on its mechanical deformations~\cite{park2003, mylvaganam2006, zhang2006}, which plays the key role in the design of CNT-based gas sensors.~\cite{kong2000}.

Among the radially deformed SWCNTs, the fully collapsed structure with two strained edges bridged by a flat middle section~\cite{yakobson1996,yakobson1996PRL,chang2010,shklyaev2011} attracts intense interest because of its physical and chemical properties~\cite{liu2012, martel1998,lammert2000} associated with its flat ribbon-like middle part~\cite{lammert2000}. However, the fully collapsed SWCNTs are stabilized by van der Walls (vdW) interaction between two opposing flat walls (with typical interlayer distance $d_0\approx3.4\,\mathrm{\AA}$), leads to irreversible collapsing. The vdW interaction may also induces twist and bending of the flat section in a fully collapsed SWCNT. Moreover, the edge section of a collapsed SWCNT is highly strained. As it is widely known, with the increase of hydrostatic pressure, a cylindrical SWCNT at first becomes oval-shaped, then becomes peanut-shaped.~\cite{zang2004} Between these two shapes, there exists a critical shape under certain pressure which is the unstrained flattest configuration of SWCNTs. The radial deformed SWCNT under the applied hydrostatic pressure with a flat section similar to that of fully collapsed SWCNTs have advantages over fully-collapsed ones:  1) it can be shifted back to the state with a circular cross section reversibly; 2) there is no twist in flat section since vdW interaction can be neglected; 3) there is no strain in the two edges, therefore avoids the strain-induced change of electronic band structure.

In the present manuscript, we will theoretically determine the shapes of the unstrain flattest SWCNTs. Although the problem has been studied by molecular dynamical simulations and numerical calculations,~\cite{zang2004}  there still lack the analytical explicit expressions which can provide a design principle for CNT-based devices. We will accomplish this task by carefully studying the transition of the SWCNT deforming from a convex shape to concave shape, give an analytical expression of the critical shape for an unstrained flattest tube. Based on this analytical expression, we can calculate critical pressure for the unstrained flattest SWCNTs, as well as the orbital hybridization of carbon atoms. The latter is important for studying the absorption of molecules on SWCNTs.

The equilibrium shape of deformed SWCNTs is determined by the minimization of free energy under certain constraints.
In the present problem, the free energy contains the elastic energy and the pressure term, $F_{b}=E_{b}-p\,\Delta V$. Although both bond bending and bond stretching contribute to the elastic energy of SWCNTs, at the energy scale $\Delta p\cdot V_0\sim 1\mathrm{eV}\approx k_c$, ($k_c$ is the bending modulus of SWCNTs~\cite{tersoff1994,mu2009}), only the bond bending effect matters.~\cite{oy1997} The bond-bending energy of a SWCNT can be described by its curvatures~\cite{yakobson1996,oy1997},
\be\label{bending}
E_{b}=\frac{k_{c}}{2}\oint\,(2H)^{2}\, dA+\bar{k}\oint\, K\, dA,
\ee
Here $H$ and $K$ are the mean curvature and Gaussian curvature of the
surfaces of carbon atoms, and $k_{c}=1.17\,\mathrm{eV}$~\cite{oy1997} is consistent with the result of Tersoff {ie et al.}~\cite{tersoff1994}.
The expression of free energy can be mapped to a $2$D one. Consider a straight SWCNT with radius $\rho(\phi)$, without the inclusion of its two end-caps, the surface of the tube can be described in cylindrical coordinates as, $\vec{r}(\phi,\, l)=
{\rho(\phi)\,\cos\phi,\,\rho(\phi)\,\sin\phi,\, l}$. Here, $0<\phi<2\pi,$ and $0<L<L_{0}$, with $L_{0}$ the length of straight
tube axis. The surface's mean and Gaussian curvature are $2H=-\left(\rho^{2}+2\rho'^{2}-\rho\rho''\right)\left(\rho^{2}+\rho'^{2}\right)^{3/2},\; K=0$. Comparing with the relative curvature $k_{r}=|\left(\rho^{2}+2\rho'^{2}-\rho\rho''\right)|/\left(\rho^{2}+\rho'^{2}\right)^
{3/2}$ of a plane curve $\rho=\rho(\phi)$, the bending energy can be rewritten as $E_{b}=k_{c}L_{0}/2\oint\, k_{r}^{2} ds$, where $s$ the arc parameter of boundary curve $\mathcal{C}:\rho(\phi)$, and line element
$\mathrm{d}s=\left[\rho^{2}+\left(\mathrm{d}\rho/\mathrm{d}\phi\right)^{2}\right]^
{1/2}\mathrm{d}\phi$.

The equilibrium shape of a deformed SWCNT is the solution of the $2$D variation problem $\delta^{(1)} F=0,\,\delta^{(2)} F>0$, with
\be\label{objectfunction}
F=\frac{k_{c}}{2}\oint k^2_{r}(s)\mathrm{d}s+ p\int\mathrm{d}A+\lambda\left[l_0-\oint\mathrm{d}s\right]
\ee
where, $\lambda$ is the Lagrange multiplier, which is introduced to keep the circumference of tube at constant $l_0$.
Distorting the curve $\mathcal{C}$ by a small perturbation $\psi(s)$ along the normal direction of the curve, the variations in $\delta^{(1)} F$ are, 
\ba\label{3variations}
\delta^{(1)}\oint\mathrm{d}s&=&-\oint \mathrm{d}s\, k(s)\psi(s),\nonumber\\
\delta^{(1)}\oint\mathrm{d}s\, k^2(s)&=& \oint \mathrm{d}s\, \left[k^3(s)+2k''(s)\right]\psi(s),\\
\delta^{(1)}\oint \mathrm{d}A&=& \oint \mathrm{d}s\, \psi(s).\nonumber
\ea
Then, equation of the equilibrium shape of SWCNTs can be obtained,
\begin{equation}\label{eq:nonlineardifferentialequation}
p\,+\frac{k_{c}}{2}\,k_{r}(s)^{\,3}+k_{c}k_{r}''(s)+\lambda k_r(s)=0.
\end{equation}
Obviously, there is a special solution corresponding to a circular cross-sectional tube, $k_{r}=1/\rho_{0}$, which implies the necessary condition for maintaining SWCNT's circular cross section. The initial equation of Eq. (\ref{eq:nonlineardifferentialequation}),
\be\label{initial}
p\,k_{r}+\frac{k_{c}}{8}\,k_{r}^{\,4}+\frac{k_{c}}{2}\left(k_{r}'
\right)^{2}+\frac{\lambda}{2}k^{2}_r=c_{1},
\ee
leads to the equation of $k_r$,
\be\label{eq:initialintegral}
k_{r}'=\pm\sqrt{c_{2}+\alpha\,k_{r}-\beta\,k_{r}^{\,4}+\gamma
k_r^2},
\ee\
with $\alpha\equiv-2 p/k_{c},\,\beta=1/4,\gamma\equiv-\lambda/k_c$.

\begin{figure}[h]
\includegraphics[width=90mm]{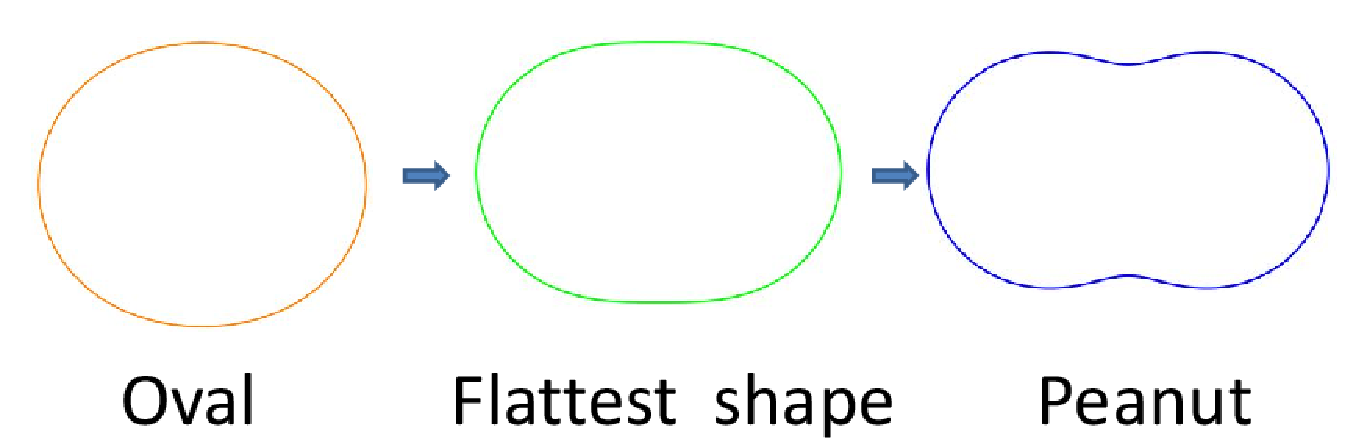}
\caption{ Illustration of shape transitions for a cross section of SWCNT with $C_{2v}$ symmetry. The critical shape for the transition from the oval shape to the peanut-like shape is the unstrained flattest shape of SWCNT.}
\label{fig1}
\end{figure}

\begin{figure}
\includegraphics[width=60mm]{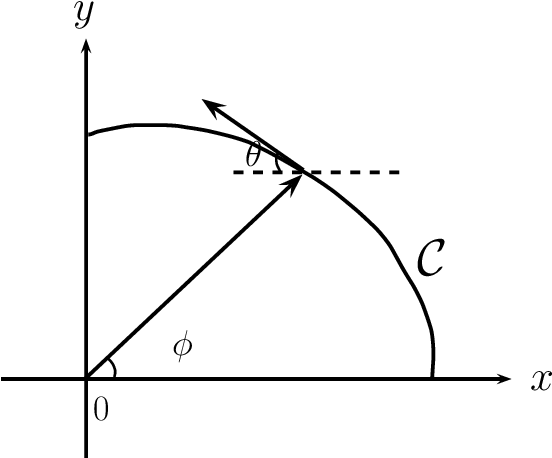}
\caption{ Sketch illustrate of the first quarter of the deformed cross section of a SWCNT with $C_{2v}$ symmetry. The $\theta$ is the angle between $x$-axis and the direction of the curve $\mathcal{C}$'s tangential line. The $\phi$ is the polar angle.}
\label{fig2}
\end{figure}

Under hydrostatic pressure, the circular cross-sectional tube can deform to a convex structure with lower symmetry $C_{nv}$. With the increase of pressure, the curvature changes continually from $k_{r}>0$ (convex shape) to $k_r<0$ (concave shape). A critical shape for the transition from convex to concave satisfies $k_{r}\geq 0$, with the minimal curvature being exactly zero, $k_{r,\mathrm{min}}=0$. The critical shape is the flattest shape for unstrained SWCNTs.  In particular, for the deformed SWCNT with $C_{2v}$ symmetry, the convex shape is "oval" shape, and concave shape is "peanut-like" shape, as shown in Fig.~\ref{fig1}. Equation (\ref{eq:initialintegral}) can be solved numerically by iteration~\cite{zang2004}. Exploiting the symmetry of deformed shape, only the first quarter of curve $\mathcal{C}$ is necessary to be considered, as shown in Fig.~\ref{fig2}. For convenience, the length of curve $\mathcal{C}$ is re-scaled to $2\pi$. $k_r(s=\pi/2)=0$ and the fixed circumference $l_0=2\pi$ provide a constraint,
\be\label{kr2pi}
\frac{\pi}{2}=\int_{k_{r,max}}^{0}\,\frac{\mathrm{d}k_r}{\sqrt{c_{2}+\alpha\,k_{r}-\beta\,k_{r}^{\,4}+\gamma
k_r^2}},
\ee

The direction of tangential line changes from $\theta=\pi/2$ to $\theta=0$, as the arc parameter $s$ changes from $0$ to $\pi/2$, which provides another independent equation,
\be\label{krtheta}
-\frac{\pi}{2}=\int_{k_{r,max}}^{0}\,\frac{k_r\mathrm{d}k_r}{\sqrt{c_{2}+\alpha\,k_{r}-\beta\,k_{r}^{\,4}+\gamma
k_r^2}}.
\ee
Equations (\ref{kr2pi}) and (\ref{krtheta}) can be calculated self-consistently. Here, we have already used relation $\mathrm{d}\theta=k_r \mathrm{d}s$ in the derivate of Eq.~(\ref{krtheta}).
In principle, the flattest shape with high order symmetries ($C_{nv},\,n\geq 3$ ) can be obtained similarly.

In the application of deformed SWCNT related devices, such as a gas sensor, it is necessary to derive an explicit expression of the critical pressure for the flattest shape, in order to obtain the design rule for working parameters. In this manuscript, we will give an approximate analytical solution of Eqs.~(\ref{kr2pi}) and (\ref{krtheta}).

As known from the exact numerical solution of $k_r$, 
$c_0$ is very small, and $\gamma=0$ is a good initial guess for self-consistent solution of shape equation Eq.~(\ref{kr2pi}) and (\ref{krtheta}). It is natural to get an approximate solution with $\gamma=0$, then compare it with the numerical solution to check the consistency.
For $\gamma=0$, $k_{r,\mathrm{min}}=0$, Eq.~(\ref{eq:initialintegral}) implies $c_2=0$, it can be reduced to
\be\label{ceq0}
\frac{\mathrm{d}k_{r}}{\sqrt{\alpha\,k_{r}-\beta\,
k_{r}^{4}}}=\mathrm{d}s=\frac{1}{k_{r}}\,\mathrm{d}\theta,
\ee
with the additional necessary condition of equilibrium critical shape: $k_{r}$ reaches its maximum at $\theta=\pi/2$, and its minimum at $\theta=\pi/2+\pi/n$. Thus, we have, $k_{r,\, n}(\theta)=(4\alpha)^{(1/3)}\,\sin^{2/3}\left[3\left(\theta+c_{1}(n)\right)/4\right]$ and $c_{1}(n)(\theta)= 5\pi/6-\pi/n$.
The constant length of curve $\mathcal{C}$, $l_0=2\pi$ requires
\be\label{length}
\frac{l_{0}}{2n}=\frac{\pi\rho_{0}}{n}=(4\alpha)^{(-1/3)}\,\int_{\pi/2}^{\pi/2+\pi/n}\,k_{r,\, n}^{\,-1}(\theta)\,\mathrm{d}\theta\,=\frac {c_{2}(n)}{(4\alpha)^{(1/3)}}.
\ee
The constants $c_2(n)$ can be derived as,
\[
c_{2}(n)=\frac{4\sqrt{\pi}\Gamma\left(\frac{7}{6}\right)}{\Gamma\left(\frac{2}{3}\right)}-\frac{4}{3}\left[_{2}F_{1}^{}\left(\frac{1}{2},\,\frac{5}{6},\,\frac{3}{2},\,\cos^{2}\frac{3\pi}{4n}\right)\,\cos\left(\frac{3\pi}{4n}\right)\right].
\]
Here, $_2F_1(a,b,c,z)$ is the hypergeometric function~\cite{abramowitz1972}. Some constants $c_{2}(n)$ for a unstrained flattest tube with $C_{nv}$ symmetries are listed in Table.~\ref{tab1}.
\begin{table}[htbp]
\caption{\ The $c_{2}$ values for the approximate solution of a unstrained flattest SWCNT whose cross section has $C_{nv}$ symmetry.}
\label{tab1}
\begin{ruledtabular}
\begin{tabular}{lccccccccc}
$n$ & 2 & 3 & 4 & 5 & 6 & 7 & 8 & 9 & 10\\
    \hline
    $c_{2}$ & 4.324 & 3.728 & 3.372 & 3.124 & 2.936 & 2.788 & 2.665 & 2.562 & 2.473\\
\end{tabular}
\end{ruledtabular}
\end{table}
The curvature $k_r(n),\;n\geq 2$ has the form of
\be\label{krn}
k_{r,\, n}(\theta)=\left(\frac{n\, c_{2}(n)}{\pi\rho_{0}}\right)\,\sin^{2/3}\left[\frac{3}{4}\left(\theta+\frac{5\pi}{6}
-\frac{\pi}{n}\right)\right].
\ee
In particular, $c_{2}(2)=4.3244$, predicts the critical shape for the transition from oval-shaped tube to peanut-like shape.

The equilibrium shape of the unstrained flattest SWNCTs with $C_{2v}$ symmetry can be described by parametric equations,
\begin{eqnarray}
\left\{
\begin{array}{lll}
x_{2}(\theta) & = & \frac{2\pi\rho_{0}}{c_{2}}\,\sin\left(\frac{\theta}{4}+\frac{\pi}{4}\right)\sin^{1/3}\left(\frac{3\theta}{4}+\frac{\pi}{4}\right) ,\\
y_{2}(\theta) & = & \frac{2\pi\rho_{0}}{c_{2}}\,\left[\frac{\sin^{2/3}\left(\frac{\pi}{8}\right)}{\sqrt{2}}+\sin\left(\frac{\theta}{4}-\frac{\pi}{4}\right)
\sin^{1/3}\left(\frac{3\theta}{4}+\frac{\pi}{4}\right) \right] .
\end{array}
\right.
\end{eqnarray}

\begin{figure}
\includegraphics[width=80mm]{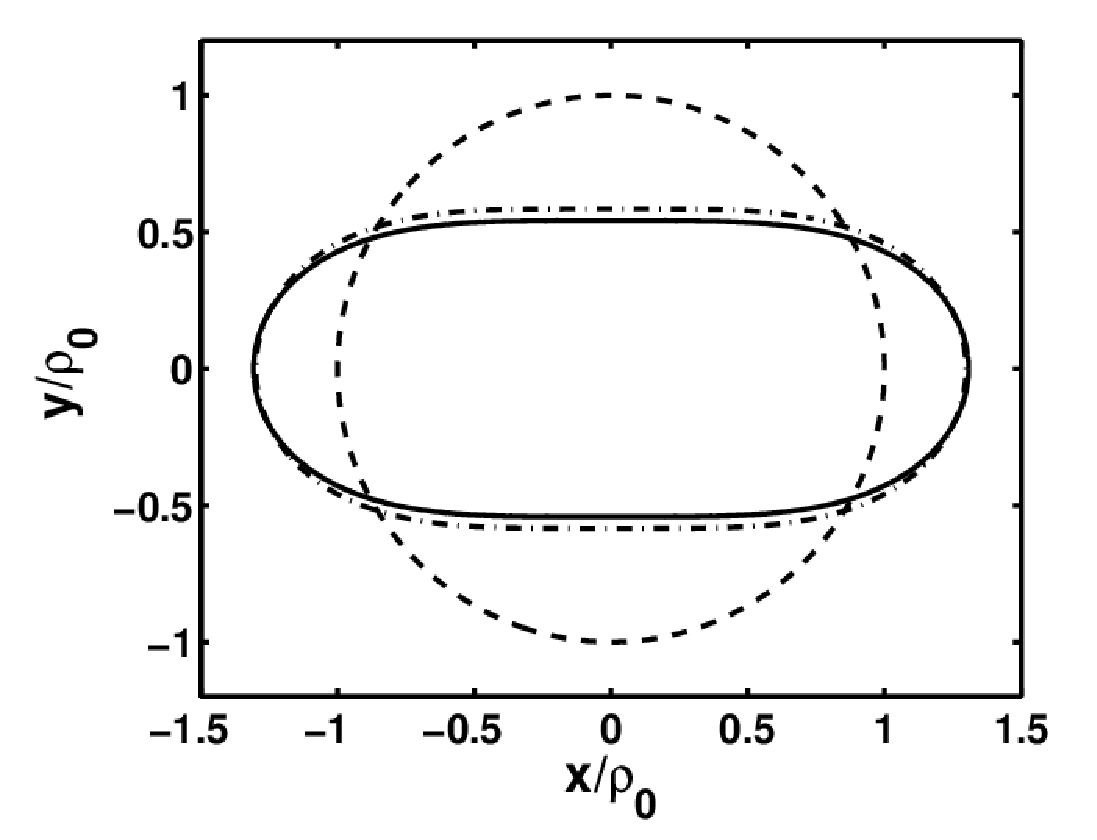}
\caption*{Comparison of the explicit solution and exact solution of  deformed SWCNTs' at critical flat shape with $C_{2v}$ symmetry. Solid line curve is the present explicit approximate solution of shape equation, dash-dot line curve is the exact numerical solution, and dashed line is unperturbed circular shape of a SWCNT. }
\label{fig3}
\end{figure}

The area of tube's cross section is
\be\label{s2}
S_{2} =  \frac{1}{2}\left(\frac{\pi\rho_{0}}{c_{2}}\right)^{2}
\int_{\pi/2}^{\pi}\,\frac{\sin\theta\sin\left(\theta/4+\pi/4\right)}{\sin^{1/3}\left(3\theta/4+\pi/4
\right)}=2.449\rho^2_0.
\ee
Compare with the undistorted tube, $S_{0}=\pi\rho_{0}^{\,2}$, the geometric constant is $\mathcal{G}=S_{2}/S_{0}=0.78$, which is in good accordance with the exact value $\mathcal{G}=0.8195$.~\cite{zang2004}

\begin{figure}
\includegraphics[width=80mm]{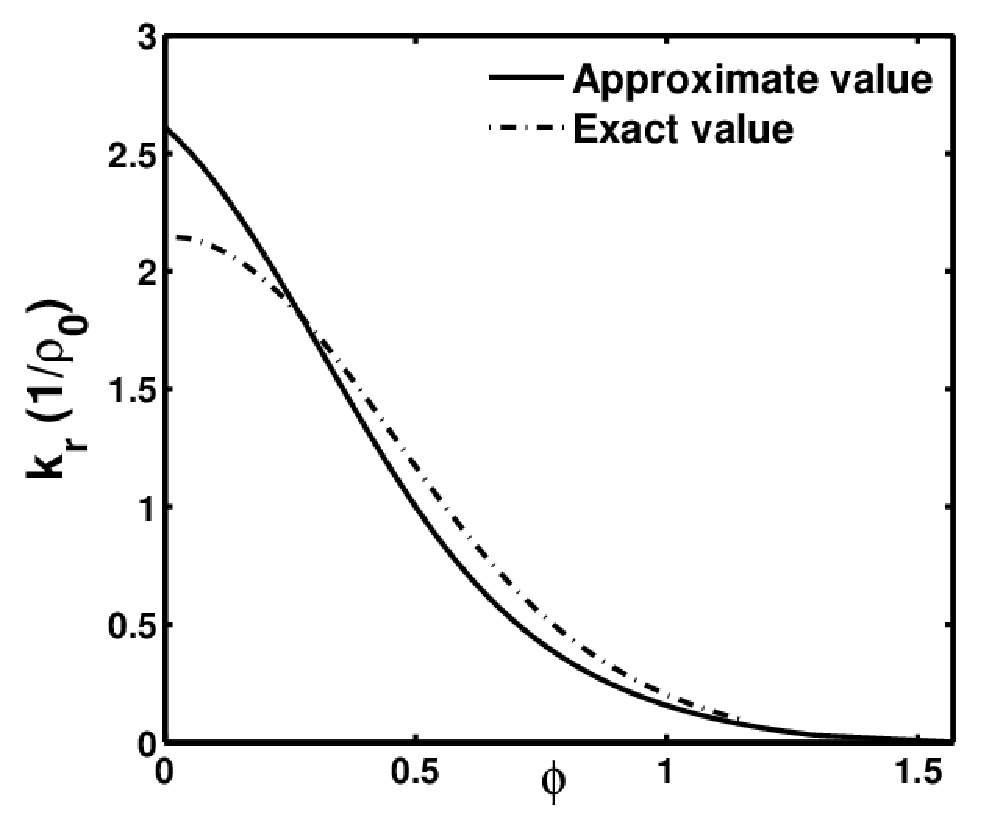}
\caption{Comparative study of the approximate and exact results of curvature distribution of unstrained SWCNTs at the transition from oval shape to peanut shape. Here $\phi$ is the polar angle shown in Fig.\ref{fig2}. The curvature $k_r$ is in unit of $1/\rho_0$,  with $\rho_0$ being the radius of unperturbed circular cross section of a SWCNT.}
\label{fig4}
\end{figure}

The curvature distribution characterizes the flatness of the deformed SWCNTs. The curvature distribution provides useful information of the bond hybridization of SWCNTs, which governs the ability of absorbing gas molecules by SWCNTs. We also perform comparative study of the curvature distribution of our approximate result and the exact solution, as shown in Fig.~\ref{fig4}. The exact curvature distribution can be calculated self-consistently according to Eqs.~(\ref{kr2pi}) and (\ref{krtheta}).
The hybrid orbital of carbon atoms at a SWCNT is sensitive to the local curvature of the tube. For a SWCNT, in the local coordinate of a given carbon atom $i$, the three neighbours of the center atom i have coordinates,~\cite{mu2009}
\ba\label{xyz}
x_{0j}	&=&	\cos\theta_{j}-\frac{(a_{0}/R)^{2}}{3}\,\sin^{4}\theta_{j}\,\cos\theta_{j},\nonumber\\
y_{0j}	&=&	\sin\theta_{j}\,+\,\frac{(a_{0}/R)^{2}}{6}\,\sin^{3}\theta_{j}\,\cos(2\theta_{j}),\quad j=1,2,3, \nonumber\\
z_{0j}	&=&	\frac{(a_{0}/R)}{2}\,\sin^{2}\theta_{j},
\ea
with $\theta_{1}=\theta_{c}$, $\theta_{2}\approx\theta_{c}+2\pi/3$, and $\theta_{2}=\theta_{c}+4\pi/3$ are angles between "bond curves" and the direction of tube axis, at the position of atom $i$.~\cite{oy1997}. Here, $a_0=1.42\AA$ is the equilibrium bond length of carbon-carbon bond, $R$ is the radius of the SWCNT, $a_0/R$ characterizes the curvature dependent properties of SWCNTs.
The distance between atom $i$ and their three neighbours can be described as $r_{ij}^{0}=1-\frac{(a_{0}/R)^{2}}{24}\,\sin^{4}\,\theta_{j}$. The "directions" (the direction of symmetric axis) for three $sp^{2}$
orbits can be written as $\hat{e}_{j}=\{x_{0j},\, x_{0j},\, z_{0j}\}/r_{ij}^{0}$.

In the local coordinate system associated with give carbon atom $i$, in general, the $\pi$ orbit and three $sp^{2}$ orbits can be decomposed as, $|\pi\rangle	=	c_{1}|2s\rangle+\sqrt{1-c_{1}^{2}}\,|2p_{z}\rangle$, and $|sp^{2}\rangle_{j}	=	c_{2}\,|2s\rangle+\sqrt{1-c_{2}^{2}}\,\left(e_{jk}|2p_{k}\rangle\right),\quad k=x,y,z$.
with, $c_{1}=\sqrt{\frac{6\alpha^{2}}{32-3\alpha}}\approx\frac{\sqrt{3}}{4}\alpha$, $c_{2}=\sqrt{\frac{32-9\alpha^{2}}{3(32-3\alpha^{2})}}\approx\frac{1}{\sqrt{3}}\left(1-\frac{3}{32}\alpha^{2}\right)$, and $\alpha\equiv a_{0}/R$.
Therefore, $|\pi(\alpha)\rangle=\frac{\sqrt{3}}{4}\alpha|2s\rangle+\left(1-\frac{3}{32}\alpha^2\right)|2p_{z}\rangle$, which plays a key role in the electronic properties and chemical activities of SWCNTs.~\cite{park2003}  For graphene, $|\pi(\alpha=0)\rangle=|2p_{z}\rangle$. By adjusting applied hydrostatic pressure, the $|\pi\rangle$ at the main part of a SWCNT can be switched between $|\pi(\alpha)\rangle$ and $|2p_z\rangle$ reversibly.

In summary, we have theoretically investigate the unstrained flattest shape of SWCNT under hydrostatic pressure, which can recover to its original circular cross-section after withdrawing the pressure. We find a good approximate solution for the shape of this flattest
SWCNT, and theoretically determine the curvature distribution, as well as the curvature-dependent hybrid orbitals of the SWCNT. The present results are in good agreement with the exact numerical solution, and it is suitable to be used to study the design rule in related CNT-based nano-electronic devices due to its analytical form. Our approach can be generalized to investigate other inorganic and organic elastic membrane systems~\cite{mu2012,mu2013}, including the self-assembled polymer materials and colloidal aggregations.~\cite{wu2005,wu2006}

W. Mu and J. Cao acknowledge the  acknowledge the financial assistance of Singapore-MIT Alliance for Research
and Technology (SMART), National Science Foundation (NSF CHE-112825). W. Mu and Z-c. Ou-Yang acknowledge the supported of National Science Foundation of China (NSFC) (Grants No. 11074259 and 11374310), and the Major Research Plan of the National Natural Science Foundation of China (Grant No. 91027045).  J. Cao has been partly supported by the Center for Excitonics, an Energy Frontier Research Center funded by the U.S. Department of Energy, Office of Science, Office of Basic Energy Sciences under Award No. DE-SC0001088.

\providecommand{\noopsort}[1]{}\providecommand{\singleletter}[1]{#1}%

\end{document}